# The distribution and lifetime of powerful radio galaxies as a function of environment and redshift

David Garofalo[1], Chandra B. Singh[2], Alexa Zack[1]

1. Department of Physics, Kennesaw State University, Marietta GA 30060, dgarofal@kennesaw.edu; azack1@students.kennesaw.edu
2. The Raymond and Beverly Sackler School of Physics and Astronomy, Tel Aviv University, Tel Aviv 69978, Israel. chandratalk@gmail.com

## Abstract

Correlations between jet power and active time for z < 0.1 high excitation and low excitation radio galaxies are explored as well as evidence in favor of a specific, non-random distribution for these objects including mid-infrared emitting radio galaxies as a function of environment and redshift. In addition, so-called weak line radio galaxies with FRII jet morphology have been identified as a class of active galaxies in the process of shutting down. This paper identifies common features between these seemingly disparate phenomena described above for the population of radio galaxies, and strings them together by way of a simple phenomenological framework that has shed light on the radio loud/radio quiet dichotomy, the jet-disk connection, and the distribution of all active galaxies as a function of redshift.

I. Introduction

Despite the emergence of widely accepted models for black hole accretion, jet formation, and AGN feedback over the past few decades, observation and theory remain discrepant. Among the most fundamental open questions are (1) the non-random redshift distribution of AGNs with and without powerful jets [4],[13],[15],[18],[19],[47],[51],[52],[55], and the nature of the interplay between black hole engines and environment behind the generation of strong and weak jets [7],[9],[17], [24], [31], [43], [48]. High excitation radio galaxies (HERGs) are overwhelmingly associated with powerful Fanaroff-Riley type II jets (FRII jets [17]) and their occurrence is greater at redshift of about 2 and decreases above and below that. Radio galaxies with FRI jet morphology [17] are overwhelmingly associated with low excitation accretion states - low excitation radio galaxies (LERGs) - and their numbers peak at redshifts below that of the HERGs (see [63] Figure 13 for the redshift distribution of HERGs and LERGs).

The idea most subscribed to for the generation of powerful jets from black holes is the spin paradigm [10],[40],[50],[59],[64] whereby high spinning black holes surrounded by prograde rotating accretion disks give rise to powerful jets. Although generally in the narrower context of thick disks, this idea has received support from numerical simulations over the last two decades [27],[35],[59],[60]. This combination of analytic and numerical work spanning four decades has culminated in a picture for powerful jet formation associated with thick disk, advection dominated accretion, around high prograde spinning black holes.

Combining the observational fact that major mergers peak around z=2 with the idea that mergers trigger AGN activity, leads to an expectation of a peak for powerful jet-producing AGN around z=2. But observations of a significant difference in the peak distribution of jetted AGN with FRII and FRI morphologies forces models beyond this simple one-to-one relationship between mergers and AGN activity. Because most AGN are radio quiet or with weak jets, spin paradigm models - which are grounded in the idea that accretion is mainly chaotic so as to mostly spin black holes down – rely on mergers to produce the needed high spins. This forces an assumption about mergers that is not observed, namely an additional pocket of mergers triggering AGN at lower redshift (z~1) that leads to powerful FRI radio galaxies accreting in hot mode. The absence of merger signatures associated with this population is well documented [3],[5],[14],[16],[24][28],[30]. In addition, the distribution of high black hole spin predicted by numerical simulations is not sufficient to explain the observed distribution of FRII quasars and FRI radio galaxies. The constraints on powerful jet formation from the spin paradigm must therefore be relaxed so as to allow powerful jets with black hole spin as low as about 0.8 [18],[19].

Further constraints on these AGN come from estimates of timescales for FRII radio galaxies which appear to experience order of magnitude smaller active phases than FRI radio galaxies [45]. In addition to confirming this trend, [61] identify powerful restrictions on models stemming from a correlation between jet power and active time in HERGs and LERGs. Despite a large scatter, and in a model-dependent way, not only is the active time for HERGs limited to about $10^7$ years, the data is compatible with an inverse correlation between jet power and active time. In the LERG population, on the other hand, the active time is an order of magnitude or more greater and the correlation between jet power and active time appears, or is compatible with being, direct (see Figure 1).

In addition to jet power/active lifetime correlations, HERGs and LERGs distribute themselves unevenly among isolated field galaxies and cluster groups, with LERGs dominating in the latter (see [23] for a recent review). For the HERGs that do appear in clusters, however, they appear to prefer higher redshifts [36]. Additional clues to the nature of these radio galaxies come from radio morphology, with weak line radio galaxies with FRII morphology being identified as AGN that are shutting down and therefore associated with intermittent accretion [57]. Finally, so-called 'mid-infrared emitting AGN', dominated by mid-infrared emission possibly attributed to the existence of a torus structure supplying the accretion material, are similarly observed to be non-randomly distributed in different environments and possibly redshift [39].

In view of this multitude of recent and apparently disparate set of observations, models that can more simply incorporate mergers, accretion, their connection to powerful jet production and their relation to very different active time, that can also explain the particular combinations of observed environment and jet morphology, are needed. In this paper we show how to understand the seemingly very different aspects of the AGN population described above in a unified way from our phenomenological framework, the so-called gap paradigm for black hole accretion and jet formation [20]. At the heart of this paradigm is the triggering of the most powerful FRII quasar jets in the aftermath of mergers onto retrograde rotating black holes (a small subset of merger-triggered AGN) that must first spin down and then up into the prograde direction due to prolonged accretion (see [41] and [42] for variations on this theme). In Section 2 we highlight elements of the phenomenological model. In Section 3 we apply it to the connection between observed jet power and active lifetime. In Section 4 we apply it to observations of cluster and field radio galaxies and mid-infrared AGN distribution. In Section 5 we apply it to the WLRG/FRII objects. And in Section 6 we summarize and conclude.

2. The Gap paradigm

The gap paradigm for black hole accretion and jet formation is a scale-invariant phenomenological framework for understanding the redshift distribution of AGN [20] that we here briefly review. In the aftermath of a merger that funnels cold gas toward the newly formed rotating black hole, the accretion disk has a 50% chance of aligning or anti-aligning its angular momentum vector with that of the black hole as a result of the Bardeen-Petterson effect. However, because retrograde accretion is an unstable accretion configuration [34],[46], only initially retrograde configurations with small ratio of disk mass to black hole mass will remain in retrograde mode so the fraction of retrograde accreting black holes remains a small fraction of the entire accreting black hole population. In addition, prolonged accretion forces an evolution away from retrograde accretion configurations by spinning the black hole down toward zero spin and then up into the prograde regime. And because the spin-down timescale during the retrograde phase is short-lived, even the few objects that experience retrograde accretion rapidly evolve away from such states, further contributing to making retrograde configurations the overwhelming minority among accreting black holes. Whereas retrograde accreting configurations are associated with the presence of a jet, this is only true in prograde regimes for particular values of black hole spin and state of accretion. These ideas give the reader a sense of how to understand the paucity of jetted AGN among the overall AGN class, the observational fact of the so-called radio loud/radio quiet dichotomy. Those details, however, appear elsewhere and are not the focus of this work. Nonetheless, the same ideas that allow that understanding will shed light on the observational evidence discussed above. To see how, Figure 2 illustrates the time evolution of an originally retrograde accretion configuration under three different initial conditions for jet power. The quantitative prescription for jet power is given as

$$L_{jet} = 5 \times 10^{47} \text{ erg s}^{-1} f(a) (B_d/10^5 G)^2 m_9^2 \quad (1)$$

where $f(a)$ is a spin-dependent function - such that $-1 < a < 1$ - that comes from the numerical solution to the equations for the combined Blandford-Znajek and Blandford-Payne contributions to jet power, $B_d$ is the strength of the magnetic field threading the inner accretion disk in Gauss and $m_9$ is the black hole mass in units of 1 billion solar masses. Our focus is on the parameter space that allows for weak/strong jets via equation (1). Although near-zero black hole spin systems produce negligible jet power, intermediate to high spin values may produce jet powers that although greater in the retrograde configurations, are within an order of magnitude of each other. In order to model the measured jet powers in extragalactic radio sources, magnetic fields in the inner disk of a few thousand Gauss are needed. Uncertainty, however, plagues our understanding of the origin of these inner magnetic fields [6],[29],[37],[38]. The dependence of jet power on black hole mass is crucial for understanding our applications in this work. The three panels in Figure 2, in fact, can be thought of as distinguishing each other by the black hole mass and will be described in greater detail in the application sections. It is worth noting, however, that super-Eddington accretion is not a fundamental feature of the model, at least insofar as jet feedback is concerned.

3.  Active lifetimes versus jet power in high and low excitation radio galaxies

In this section we explore the aforementioned model-dependent observations in [61], from the perspective of the gap model. We appeal to the three panels in Figure 2 to describe the qualitative evolution of initially radiatively efficient thin disk accreting retrograde black holes. The differences in Figure 2 a, b, and c, are grounded in the jet power produced in the initial retrograde configuration. Because accretion in retrograde mode must spin the black hole down to zero spin, and continued accretion will subsequently spin the black hole up in the prograde direction, there is a natural spin evolution in the paradigm that cannot be altered. In Figure 2a, we have the weakest jet power. Such an accreting black hole is relatively weaker in its jet feedback on the accretion state because it fails to effectively heat the ISM (see [20] for details and references on how this comes about in the model), and as a result no hot ISM gas that can eventually accrete back onto the black hole occurs and the radiative efficiency of the disk remains unaltered. Such black holes accrete close to the Eddington rate which means a scale-invariant timescale to spin down to zero spin of about $8 \times 10^6$ years (e.g. [33]). Because prograde accreting black holes in thin disk configurations constitute weaker jet producers, these Figure 2a objects quickly evolve into radio quiet or jetless AGN/quasars. In Figure 2b we show the time evolution of an initially retrograde accreting thin disk whose jet power is larger compared to that of Figure 2a. Here, jet feedback is capable of heating the ISM which eventually accretes onto the black hole as hot Bondi accretion flow, or in ADAF configuration. The effectiveness of jet feedback in altering the mode of accretion is, however, relatively mild and slow in taking effect, resulting in a timescale greater than about $10^7$ years as a result of the fact that Eddington-limited accretion tends to dominate during the entire retrograde phase and ADAF mode accretion is not present until the system accretes in prograde mode. In Figure 2c, instead, we see the evolution of an initially retrograde thin disk that produces a powerful jet whose feedback relatively rapidly affects the ISM and the mode of accretion. Hence, an ADAF accreting black hole is formed while still in retrograde mode. This means the system has transitioned away from being a HERG on a timescale that is less than about $10^7$ years. Of crucial importance is the fact that ADAF accretion slows down the accretion timescale by at least a factor of two. In fact, the cutoff between thin disk and ADAF mode accretion is theoretically placed at 0.01 times the Eddington accretion rate. At a qualitative level, therefore, Figure 2 allows one to appreciate how LERG states (ADAFs) live orders of magnitude longer than HERGs. The timescale for HERGs and LERGs, therefore, constitutes our first phenomenological constraint: Retrograde accreting black holes accreting at the Eddington limit are spun down in $8 \times 10^6$ years, which constitutes an upper limit for such HERG timescales because the possible evolution of the accretion state into an ADAF on a shorter timescale implies that such HERGs live even less. If - as seen in Figure 2a - the radiatively efficient state is prolonged into the prograde spin regime, the system evolves away from being a HERG as a result of the jet quenching or suppression. Therefore, for all the possible retrograde states that might emerge in the paradigm, Figure 2 is sufficient in illustrating the idea that in the paradigm, HERG lifetimes are comparatively well-constrained to no greater than about 10 million years. These timescales are surprisingly compatible with the results of [61] (Figure 1). The LERG states, on the other hand, have active phases that are an order of magnitude or more larger, and depend ultimately on the supply of gas.

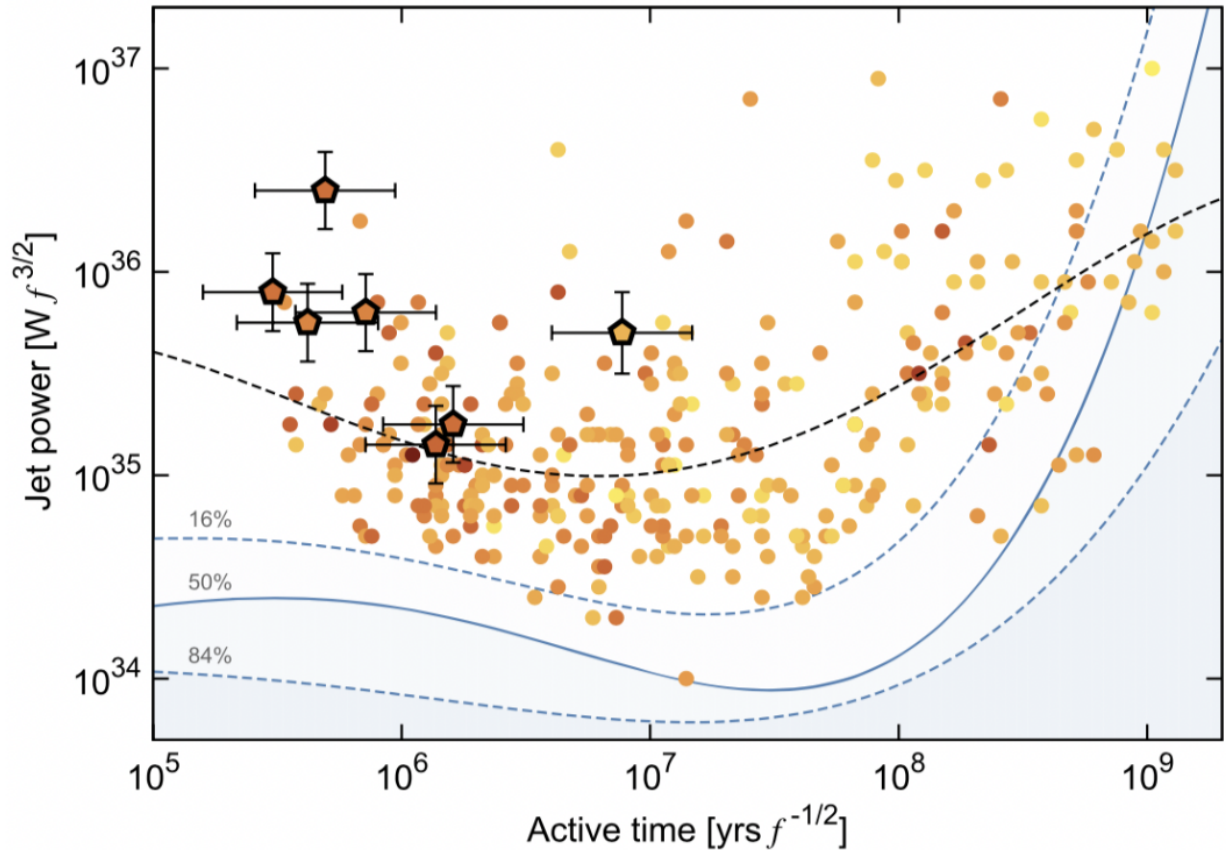

Figure 1: Jet power versus active lifetime from Figure 16 of [61]. Pentagons represent HERGs while circles are LERGs. Although the scatter is large, jet power is compatible with being inversely dependent on active time for HERGs while the opposite emerges for LERGs. The value of $f$ is 5.

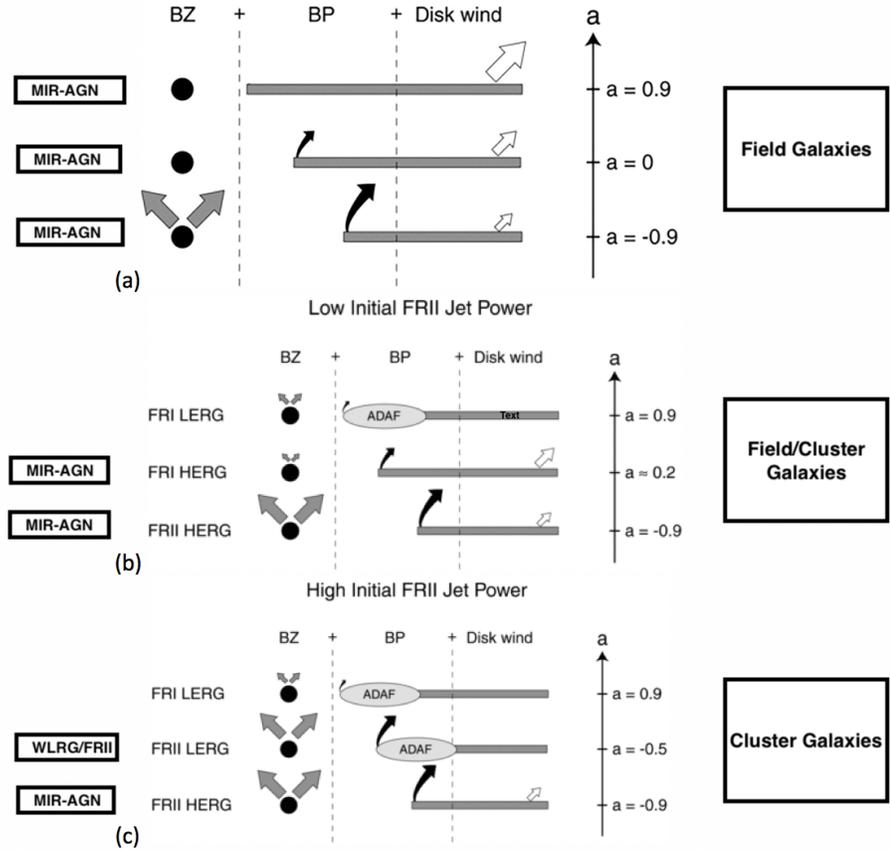

Figure 2: (a) When the FRII quasar jet is not sufficient to alter the accretion mode, the system evolves into a prograde regime while remaining in a radiatively efficient state. As a result of this continued radiative efficiency of the disk, the disk wind suppresses the jet and a radio quiet quasar/AGN emerges. (b) Time evolution of an initially retrograde accreting black hole with intermediate jet power. The radiative efficiency of the initially radiatively efficient thin disk (lower two panels) evolves slowly into an advection-dominated disk. (c) Time evolution of an initially retrograde accreting black hole with powerful jets. The radiative efficiency of the initially radiatively efficient thin disk (lower panel) evolves quickly into an advection-dominated disk, which has a radiative efficiency that is at least two orders of magnitude smaller than the Eddington value. The jet is the result of a combination of the Blandford-Znajek process (BZ[10]) and the Blandford-Payne process (BP[11]) which explains the labels in the first and second columns. In ADAF states the radiative disk wind is absent. The arrows represent the radiatively-driven disk wind which increases with disk efficiency. The dimensionless spin parameter '$a$' evolves from retrograde to zero and then up again in prograde regime due to continued accretion. Boxes on the right indicate environment. Note the association of WLRG/FRIIs with FRII LERGs. Left column describes the system in terms of its mid-infrared emitting timescale. The longer-lasting MIR emitter in (a) results from an accreting black hole whose torus structure is longer lasting. The evolution into an ADAF/LERG accretion state is associated with an absence of a torus and an enhancement in hard X-rays. This occurs slowly for (b) and quickly for (c).

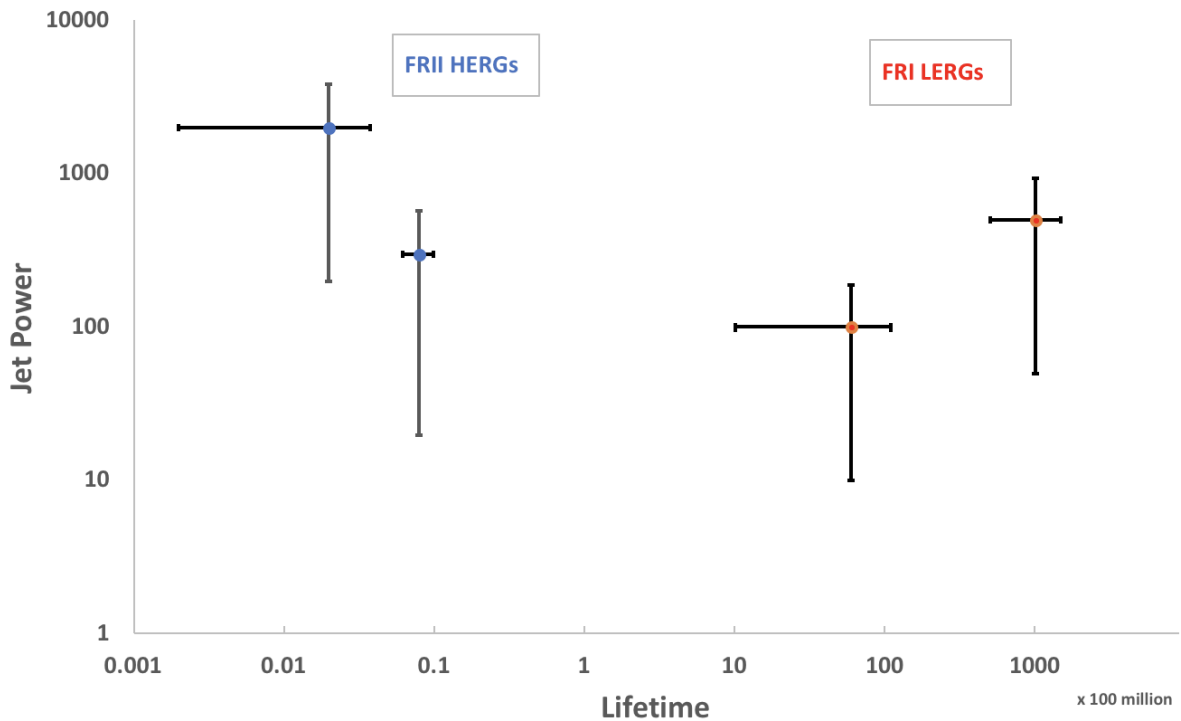

Figure 3: Jet power in units of $(B_5 m_9)^2$ where $B_5$ is the magnetic field in units of $10^5$ Gauss and $m_9$ is the black hole mass in units of $10^9$ solar masses, versus lifetime (in units of $10^8$ years) for FRII HERGs and FRI LERGs from the gap paradigm. The two FRII HERGs in blue correspond to Figure 2 paths a & b higher jet power, smaller lifetime, and path c, lower jet power, longer lifetime. The two FRI LERGs in red correspond to paths b, lower jet power and shorter lifetime and path c, higher jet power and longer lifetime. Fractional uncertainties in jet power of order unity or greater emerge just from the possible range of black hole threading magnetic field strengths. The active lifetime uncertainty for LERGs is much greater than for HERGs, since the latter are constrained by Eddington-limited accretion in retrograde mode.

In addition to timescales, [61] discovered possible correlations between jet power and active time. For HERGs they find the data to be compatible with an inverse relation between jet power and active time which we again address using Figure 2. As noticed above, the more powerful the HERG jet, the more rapid the transition to the ADAF state, and the shorter lived the HERG. If, as the middle panel in Figure 2c describes, the system transitions into an ADAF while still in retrograde mode (i.e. a =-0.5), the HERG state has lived less than the time it takes to spin the black hole down to zero spin at the Eddington limit. Hence, the lower the jet power, the greater the chance that the HERG will live as long as ~ $10^7$ years. That constitutes our simple phenomenological explanation for the inverse relation between HERG jet power and active time. The same reasons that produce an inverse relation between jet power and active time in HERGs, explain why the reverse is true for LERG states. In fact, the more powerful the HERG jet is, according to our phenomenology, the more rapidly the system transitions to a LERG state, and LERG states last longer for a given amount of gas. The weaker the original HERG jet is, the longer the system waits to become an ADAF, which means that once the LERG state sets in, less of the total available gas remains to be accreted because much of it was accreted at the Eddington limit.

In Figure 3 we construct a theoretical plot of jet powers versus lifetime for FRII HERGs and FRI LERGs, both from the perspective of the gap paradigm. The two data points in blue represent FRII HERGs which are prescribed as thin disk accretion onto high spinning retrograde black holes. The blue dot with high jet power represents paths *c* from Figure 2 which are powerful FRII quasars that rapidly evolve into ADAF accretors while still in retrograde configuration. Therefore, they live shorter time as radiatively efficient accretors or HERGs. We estimate the timescale based on Eddington limited accretion spinning the black hole down to -0.5 where the minus sign represents retrograde mode. The lower jet power blue dot corresponds to paths *a* and *b* from Figure 2 which both experience the full retrograde phase in radiatively efficient mode. As described previously, the $8 \times 10^6$ years is a precise timescale to spin the high spinning black hole down at the Eddington limit. Hence, the limit of about 10 million years and a relatively small range of possible lifetimes emerges directly from this simple Eddington-limited physics. The idea that FRII HERGs are retrograde accretors, therefore, places tight time constraints on their active time. No such tight time constraints exist for the prograde regime. The two data points in red represent FRI LERGs, which are prescribed as ADAF accretors onto black holes spinning in prograde configurations. The lower jet power red dot represents paths *b* in Figure 2 which are the late phase of an FRII HERG that is intermediate in its jet power so as not to rapidly influence the accretion state. The lifetime depends on the supply of gas but for simplicity we use the time it takes to spin the black hole up to high prograde spin at 0.01 the Eddington limit. The second red point at higher jet power represents paths *c* in Figure 2, which are FRI LERGs that originate from the most powerful FRII HERGs that therefore live longer than other FRI LERGs because their gas supply early on is fueled onto the black hole at low Eddington rates. The timescale again is that required to spin the black hole up to high prograde spin values. Hence, the lifetime for the red data should span a greater range than shown in Figure 3 but we have no way of prescribing it from the model. In other words, the constraints on timescale for the FRI LERGs depend on the amount of gas but that gas is accreting in ADAF mode which, again, theoretically is placed below 1 percent the Eddington rate. If greater jet feedback leads to ADAF accretion states that are more removed from thin disk accretion, we would predict accretion rates that are even smaller than 1% Eddington for greater jet powers. However, the values of those timescales do not lend themselves to precise prescription, unlike retrograde accretion. In fact, prograde accreting black holes remain prograde accreting black holes as long as material continues to be accreted so no simple timescale can be attributed to such states. As a result, we simply comment on the reasonableness of our expectation that LERG timescales may span a larger range than shown in Figure 3.

Despite an inverse relation between jet powers for HERGs and a direct relation between them for LERGs, equation (1) predicts a considerable scatter in jet power. This is due to the fact that jet power depends on black hole mass, black hole spin, and magnetic field, of which the latter is arguably the least constrained. In order to illustrate the effect of this on the jet power of a high-spinning retrograde accreting black hole of a certain mass, we see that the uncertainty resulting from a reasonable range of different black hole threading magnetic field values, leads to a large scatter in jet powers. An uncertainty of a factor of about 3.3 in magnetic field value leads to an order of magnitude uncertainty in jet power. These model-dependent uncertainties give us a sense of the scatter that is predicted in the model. As a result of the large scatter, in no way are we claiming that the Turner & Shabala results constitute a strong confirmation of the model. Rather, they produce an interesting compatibility that we should pay attention to in future observations, in particular when and if it becomes possible to constrain both black hole mass and nuclear magnetic field strengths.

We end this Section by commenting on the types of objects that appear in the middle panel of Figure 2b, so as not to confuse them with the FRII HERGs discussed above. The objects we are referring to, as can be seen in the figure, are instead labeled FRI HERGs. These objects possess an FRI jet morphology as a result of their prograde accretion configuration, which means a weaker Blandford-Payne contribution to jet collimation. Such objects maintain their jet as long as the spin is intermediate, somewhere in the range 0.2-0.7, beyond which the disk wind dominates the dynamics and the jet is suppressed in thin disk mode (Figure 2a top panel) or it is not suppressed due to the disk state evolving away from radiatively efficient and into ADAF mode as seen in Figure 2b top panel. These FRI HERGs or FRI quasar objects are relatively short-lived compared to the hot mode radio galaxies, the FRI LERGs. It is worth commenting, however, that spinning the black hole up in the prograde regime from about 0.2 to 0.7 requires longer than the entire retrograde spin-down. In fact, spinning the black hole up to high prograde spin from zero spin requires about $10^8$ years. Hence, the FRI HERG phase may be sustained for a decent fraction of $10^8$ years. We have chosen certain representative black hole spin values in order to simplify the evolution of the jetted AGN. However, these FRI HERGs would also appear in the intermediate prograde states of Figure 2a. Nonetheless, they are a minority among the jetted AGN population because they can only live in intermediate prograde spin states surrounded by radiatively efficient accretion. These objects are both rare in the model and in the cosmos, and have been explored in greater detail in [33]. However, the greater timescales prescribed for such objects should not be confused with the majority of HERG lifetimes which possess FRII morphology. In fact, we have consistently referred to the HERGs above as FRII HERGs. It is interesting in this context, however, to point out that the giant radio quasar 4C 74.26 [49] that appears to violate our $10^7$ year active lifetime constraint, may indeed fit into our framework as an FRI quasar.

4. Cluster versus field environments

Most HERGs and LERGs are found to have accretion rates above and below a few percent that of the Eddington accretion rate, respectively (e.g.,[7],[22],[44]). While LERGs seem to be accreting from the hot phase of the intercluster and intergalactic mediums, HERGs appear to be accreting in cold mode from mergers[24]. In luminosity-matched samples of HERGs and LERGs, the LERGs prefer the richer, cluster environments[25], [56] while HERGs tend to lie in poorer environments like field galaxies. While these trends are robust at lower redshift, they may not hold at higher redshift[23]. In this section we apply the same phenomenology that allowed us to understand the results of [61] in Section 3, to the question of why this distribution of HERGs and LERGs in cluster/field environments. In Section 4a we apply the evolution illustrated in Figure 2 to jetted AGN in field galaxies and in Section 4b we apply it to jetted AGN in cluster galaxies.

4.a. Field galaxies

Because dark matter haloes in isolated environments are smaller, the black holes in field galaxies are smaller as well. Since jet power depends on the black hole mass via equation (1), there is a direct connection between jet power and environment with field environments dominated by Figure 2a and some Figure 2b paths. For the small fraction of stable retrograde configurations in post mergers in such environments, the jet power tends to be too weak on average to affect the ISM/ICM which means accretion remains in cold mode form as seen in Figure 2a. Hence, such systems spin down to zero spin in cold mode at the Eddington rate which, as we know, requires just short of $10^7$ years. The transition around zero black hole spin means the jets turn off, but as the spin increases in the prograde direction, the jet is re-activated, although it is not as powerful as in retrograde mode (Figure 3 in [20]). For intermediate prograde spin, cold mode accretion again constitutes a HERG state, although the jet is now predicted to have FRI morphology. These are rare objects that the model explains as occurring in this narrow range of prograde spins[33]. In fact, as the spin increases beyond the intermediate prograde range, the model predicts jet suppression as the innermost stable circular orbit is close enough to the black hole horizon to make the radiative wind compete with the jet and the system is predicted to evolve into a radio quiet quasar. Overall, the model predicts little evolution in these field HERGs except for a transition through a radio quiet quasar, or radio quiet AGN phase, as the spin transitions through zero spin and back into a weaker HERG as the spin becomes intermediate, to end its evolution again as a radio quiet quasar, or radio quiet AGN, if sufficient fuel remains available. Notice that the transition through zero black hole spin in the model forces the jet power to decrease and eventually turn off. There is, therefore, an evolutionary sequence in these radio galaxies that involves continuous accretion feeding a black hole that may appear as an active galaxy that is shutting down. We will come back to this in Section 5 when we describe so-called WLRG/FRII. In addition to the weak environmental changes in field HERGs, the model predicts that field HERGs will have average jet powers that are weaker than those in clusters and active lifetimes that are longer.

4.b. Cluster environments

Whereas Figure 2a is the dominant evolutionary path for field AGN, and Figures 2b and 2c progressively less so, Figures 2b and 2c, instead, are more dominant in cluster environments as a result of the fact that such environments have larger dark matter haloes and therefore larger black holes. Thus, a good fraction of post-merger HERGs in such environments evolve rapidly away from radiatively efficient accretion states and into ADAF states while still in retrograde mode. Such objects are both retrograde and radiatively inefficient, thus FRII LERGs. The gap model, therefore, predicts both that most FRII LERGs should be observed in clusters as well as an average shorter timescale for HERGs in clusters compared with field HERGs. Both of these effects stem from the greater fraction of objects in clusters transitioning to ADAF accretion while still in retrograde mode, a feature that occurs for the more powerful jetted HERGs. Because LERGs evolve from HERGs in the paradigm, HERGs are predicted on average to live at higher redshifts than LERGs as is in fact observed [8],[26],[47]. We end this section with a brief discussion of radio quiet quasars/AGN(RQQs). Because retrograde configurations are the minority in field but also in cluster environments, RQQs are predicted by the model to be the dominant post-merger product of the funneling of cold gas into the black hole sphere of influence. RQQs will live only as long as Eddington-limited accretion continues to be supplied which is orders of magnitude less time than for LERGs. Hence, the simple evolutionary picture of Figure 2 allows us to predict/explain why lifetimes for HERGs are shortest, those for RQQs are longer, while those for LERGs are longest.

4.c. Mid-infrared emitting AGN

The thermal dust emission dominating mid-infrared wavelengths observed from active galactic nuclei is thought to originate indirectly from a torus that re-radiates disk photons. However, active galaxies whose emission in the mid-infrared is abundant, but that also produce radio jet emission, are unevenly distributed in space and time. In this Section, we apply Figure 2 to the question of why these radio, mid-infrared active galaxies dominate in field environments as opposed to clusters as found in recent observations, showing also how to understand their youth compared to active galaxies that are more dominant in X-rays. In addition, we describe why the small fraction of cluster radio and mid-infrared emitting active galaxies live at higher average redshift and experience shorter lifetimes than their field counterparts.

Again with an emphasis on black hole mass, Figure 2c describes the time evolution of the most massive black holes that begin accreting in retrograde configuration. As described, such systems contribute to a radical and rapid change in the ISM/ICM and in turn to the accretion state, which evolves into radiatively inefficient mode as hot gas is accreted onto the black hole. In such systems, the dust enshrouded emission characterizing the early stages, disappears. According to the phenomenology, they evolve away from dust enshrouded emission on timescales that are less than those characterizing Figure 2 panel (a), i.e. on timescales of a few million years. The difference between Figure 2c and Figure 2b is the slower timescale for the torus to be destroyed and for the accretion state to evolve into an ADAF which results from the weaker jet feedback compared to Figure 2c objects. Given the time evolution described above, we now apply the phenomenology to describe how these mid-infrared (MIR) emitting radio AGN should distribute themselves in space and time according to the paradigm.

MIR emitting radio AGN are thought to result from heavily obscured, post merger, dust-enshrouded AGN that also display jets [1],[2],[53],[54],[62] which are objects captured in all the lower panels of Figure 2. While the physics of the black hole accretion disk is captured in Figure 2, the torus which is thought to produce the MIR emission from the reprocessed disk radiation, is illustrated in Figure 4. Figure 2a objects have smallest black holes so their retrograde configuration fraction is smallest (smaller black holes for a given accretion mass are less stable). Since Figure 2a jets are the weakest and thus produce the least effective jet feedback compared to the other initially retrograde paths of Figure 2, their state of accretion remains unchanged. As a result, the outer regions of such systems remain as depicted in Figure 4, with the radiatively efficient disk radiation impinging on the torus structure that surrounds the disk, re-radiating in the mid-infrared. The lack of change means such systems continue to shine in the mid-infrared as long as accretion continues. In short, our phenomenology prescribes stable, longest-term, mid-infrared emission from high excitation radio galaxies in isolated field environments unlike in clusters, as we next describe.

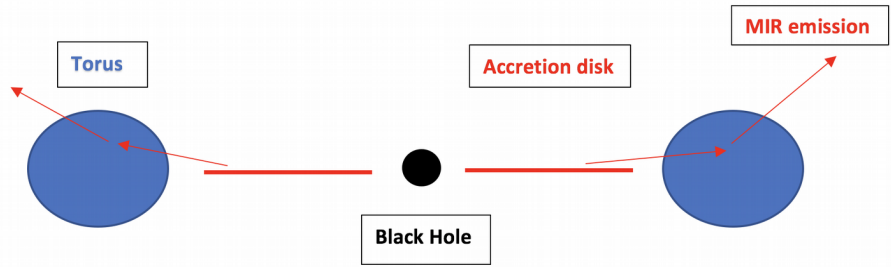

Figure 4: Mid-infrared emission from the torus, present in all three of Figure 2a panels but only in the lower two panels of Figure 1b and only the lowest of Figure 1c.

Because of the more effective jet feedback in cluster accreting black holes, they evolve away from the thin disk/torus structure of Figure 4 and into ADAFs which are spectroscopically very different from the thin disk/torus systems (Figure 5). Because the torus structure has disappeared, no reprocessing of disk radiation is possible, and no infrared radiation is generated. In fact, no big blue bump, optical/UV radiation, is produced in the first place. The inner regions are replaced by an optically thin, hot, hard X-ray emitting, plasma. Because the model predicts a transition to this ADAF accretion phase on timescales less than 10 million years for Figure 2b and 2c objects, the model prescribes MIR-emitting FRII HERGs in clusters lasting shorter timescales than in fields. In addition, such objects are also predicted to exist in such mid-infrared emitting states at higher average redshifts. In fact, the ADAF timescale is orders of magnitude longer than the radiatively efficient HERG states as a result of the fact that ADAF accretion is theoretically prescribed to kick-in at accretion rates of 1 percent the Eddington accretion rate. This means the timescale to accrete is slowed down by at least a factor of 100 and such states can supply gas for orders of magnitude longer timescales as a result. Overwhelmingly, therefore, the timescales for the MIR emitting phase in clusters is orders of magnitude smaller than the dominant hard X-ray phase. Interestingly, the observations struggle to identify MIR radio AGN in clusters compared to the field[39]. It is worth emphasizing at this point that our model prescription excludes the possibility of merger-induced hot mode accretion[23]. X-rays are a great way to find merger-induced AGN[36]. The mergers that produced the progenitor FRII HERGs occurred 10 million or more years in the past of the ADAF accretion stage. Again, the evidence points to absence of recent merger signatures in low excitation radio galaxies in clusters[3],[5],[14],[16],[24],[28],[30]. It is also important to notice that outside of the radio AGN class of objects which we have been modeling, the model prescribes a majority of radio quiet quasar/AGN objects in post mergers in both field and cluster environments. This is because the majority of post-merger cold gas accretors involves prograde accretion around spinning black holes which the model prescribes to be radio quiet quasars/AGN. This can be seen in the upper panel of Figure 2a. And such systems would retain their torus structure and therefore constitute effective MIR emitters. But they would not belong to the radio loud class of mid-infrared accretors.

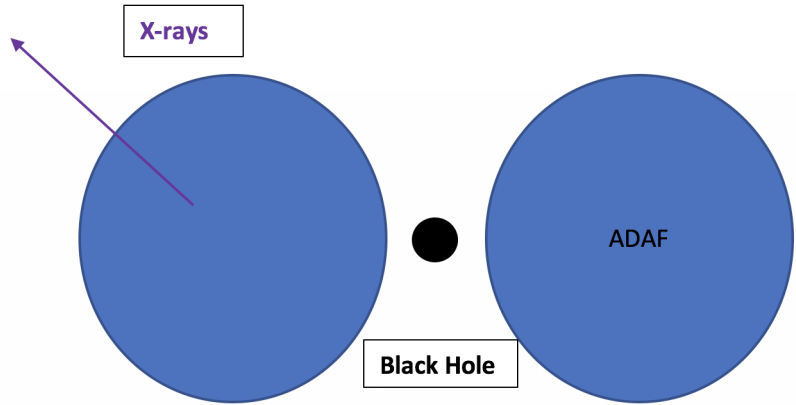

Figure 5: Hard X-ray emission from a radiatively inefficient accretion disk, characteristic of the top panel in Figure 2b and the two top panels of Figure 2c. The infrared dominated emission in the FRII HERGs of Figures 2b and 2c give way to hard X-ray emission on timescales of order millions of years.

5. WLRG/FRII

Whereas intermittent accretion onto supermassive black holes is currently a cornerstone of active galaxy evolution and feedback (See [58] for a review), it remains difficult to distinguish from continuous AGN states with low feedback. Weak line radio galaxies with FRII morphology (WLRG/FRIIs) is one class of active galaxies to which intermittent activity has been applied, in particular to PKS 0347+05[57]. Despite an abundant source of cold gas from the merger and star formation in this source, the nuclear activity appears weak while the jet morphology is that of a powerful FRII. While the authors do not rule out continued AGN activity, they prefer a picture in which the fuel to the black hole has shut off and the jet is a relic of a previous accretion phase. We explore this type of object as a class of AGN that fit into our continuous accretion picture that allows one to make precise predictions for the redshift distribution, the density of occurrence in clusters versus field galaxies, and black hole spins, for WLRG/FRIIs, constituting a range of physical parameters spanning many orders of magnitude. We argue that WLRG/FRIIs are not shutting down, but merely experiencing low accretion rates and weakening jets as they transition toward a state of renewed powerful jet feedback.

To reiterate, the basic idea behind the evolution of the state of accretion from a thin, radiatively efficient disk, to an advection dominated disk is the effectiveness of the jet feedback. If the FRII quasar/AGN (FRII HERG) is powerful, it will rapidly influence the ISM/ICM, heating it and forcing the mode of accretion to switch to hot mode on timescales of millions of years. This, again, is illustrated in Figures 2b and 2c where the FRII accretion state evolves into an ADAF from the original thin disk. But the objects we are interested in appear only in Figure 2c because the rapid transition to an ADAF means the black hole spin is in retrograde mode compared to the angular momentum of the accretion disk. Hence, we have both an FRII jet (retrograde accretion) surrounded by an accretion disk that produces weak lines (ADAF). The spin value of minus 0.5 has been chosen for simplicity. The point, however, is that the black hole spin is decreasing because of the retrograde mode and heading toward zero black hole spin as a result of continuous accretion. A zero-spin black hole has no jet which means the FRII LERG is moving toward a jetless state so it is in the process of shutting off its jet. Unlike in [57], however, the shutting off of the jet in our framework is not the result of an end of the supply of accreting material. While an Eddington-limited accreting black hole would spin down in retrograde mode in under 10 million years, and all the way up to high spin in a prograde state in about 100 million years, the transition to an ADAF makes the active galaxy evolve much more slowly. The jet shuts off in principle only at zero spin and therefore decreases continuously in retrograde mode and increases smoothly in prograde mode. However, the prograde regime needs to spin the black hole up to about a value of 0.2 before the jet becomes powerful enough to constitute an FRI radio galaxy. If this takes 10 million years or so at the Eddington rate, it will take an order of magnitude or more as it shifts toward ADAF mode where the transition is theoretically at 1% of the Eddington value or less. Hence, WLRG/FRII objects are in the process of reducing their jet feedback, and remaining in a state that is similar to LINERs, with weak lines due to the ADAF accretion mode.

A number of predictions drop out of our model. First, the fact that WLRG/FRIIs constitute in-between states whose progenitors are FRII HERGs and whose late stage are FRI LERGs, implies that they will on average display intermediate redshift distributions. Second, WLRG/FRIIs emerge from FRII quasars with the most effective jet feedback. As a result of the fact that jet power depends on black hole spin, magnetic field, and black hole mass, in the paradigm, such objects will emerge where the most massive black holes are fed. This means that such objects are predicted to dominate in rich cluster environments as opposed to isolated field galaxies [23], [47]. Third, the accretion configuration in such objects is retrograde according to the paradigm but the absolute value of the black hole spin cannot be high. The model, therefore, predicts the existence of a number of intermediate redshift weak line radio galaxies with weak or no visible jets around near Schwarzschild black holes in cluster environments. It is also worth emphasizing that only the FRII LERGs whose spin value is closer to zero constitute the objects explored by Tadhunter et al in the gap model. The FRII LERGs whose spin is intermediate as seen in Figure 2c, are in fact quite removed from shutting down their powerful jets. Hence, they are not the WLRG/FRIIs of Tadhunter's study, which instead are modeled as having an absolute value of spin that is low enough and that is approaching zero spin in retrograde mode.

6. Predictions, summary, conclusions

As we have seen, a variety of testable predictions emerge from the diagrams of Figure 2. Here, we explore how such predictions allow the model to be distinguished from the standard spin paradigm where retrograde accretion plays no significant role in jet production.

We begin by pointing out that If it is indeed the case that cluster environments tend to accrete hot gas from the outset while field galaxies tend to accrete cold gas[23], differences in radio morphology requires an explanation. Differences between FRII LERGs and FRI LERGs, in other words, should be traceable to environmental conditions of some sort such as density and pressure in the accreting hot halo gas which should vary randomly with environment and redshift. Instead, observationally we find FRII LERGs display an average redshift that is larger than the average redshift for FRI LERGs. Why would FRII jets prefer younger cluster environments? This issue appears unresolved in the spin paradigm. The gap model, on the other hand, requires it as a consequence of its time dependence as seen in Figure 2. In short, whereas the spin paradigm model at best ignores the question, the gap paradigm predicts that FRII/FRI jet morphology appear with a redshift distribution as a result of the inevitable time evolution of black hole spin due to accretion.

In addition to the issue of jet morphology, observational evidence points to larger black hole masses in FRI radio galaxies compared to the rest of the AGN family. Whereas this can be accomodated in the spin paradigm in that cluster environments possess the largest dark matter haloes, there is no immediate or natural explanation as to why FRII LERGs in clusters would have lower average black hole masses compared to FRI LERGs. This observation constitutes the coin flip-side discussed above in relation to jet morphology versus redshift. A prediction of difference in average black hole mass emerges naturally from the gap model as seen in Figure 2 as a result of the fact that FRII LERGs evolve away from such states and eventually into FRI LERGs as they continuously accrete (Figure 2c). Hence, FRII LERGs become FRI LERGs as their black holes become larger. Because non-negligible prograde spin is required to produce FRI jet morphology, it is possible to estimate the black hole mass of the offspring FRI LERG which is about 1.5 - 2 times the mass of its progenitor FRII HERG black hole. As far as we know, no attempt has been made within the spin paradigm to address differences in black hole mass between these two groups.

Beyond jet morphology and black hole mass, environmental differences are coming to light that require explanation. Cluster environments are predicted by the gap model to produce FRII HERGs at higher average redshift compared to FRI LERGs for which there is some evidence as pointed out above. Outside of the gap model, this is difficult to explain. Why would higher redshift cluster environments allow for cold gas accretion? One could argue that the higher redshift universe has a greater supply of cold gas and that some clusters are less prone to heating that gas. But since cold gas accreting black holes grow faster than low Eddington accreting black holes, tension emerges within this context with the observational fact that FRII quasars have the smallest average black hole masses among the radio loud class. Between the cold mode accreting and the hot mode accreting black hole, it is the former that grows its black hole more rapidly. Yet, observationally, the objects that possess the most massive black holes are invariably the ones accreting at the lowest Eddington rates. In the gap model, on the other hand, smaller average black hole masses in FRII quasars in high redshift clusters is a requirement as such objects are the progenitors to both the FRII/FRI LERGs. A prediction of the gap model is that overall the numbers of FRII/FRI LERGs must be less than the number of FRII quasars.

The active lifetime of a radio galaxy in high versus low excitation accretion states from the perspective of the spin paradigm cannot accommodate an inverse relation between jet power and active lifetime. Of course, the data do not unequivocally require that. On the other hand, the order(s) of magnitude difference in active timescales can be understood in the spin paradigm as the result of cold mode accretion lasting order(s) of magnitude less than in hot mode accretion. While that observation can be accomodated in the model, the specific $10^7$ year-limit that emerges from the Turner & Shabala data does not lend itself to any natural interpretation in the spin paradigm because HERGs are modeled as prograde-accreting black holes in that framework. In the gap paradigm, on the other hand, that timescale limit emerges naturally from the Eddington-limited spin-down timescale in retrograde accretion configurations.

We have shown how the phenomenology of the gap paradigm for black hole accretion and jet formation predicts an inverse relation between HERG jet power and active time and a direct relation between LERG jet power and active time albeit with a fair degree of scatter. This makes contact with observations at redshifts less than 0.1 in [61]. Despite the simplicity and qualitative character of the applied phenomenology, quantitative statements about HERGs and LERGs can be made that are surprisingly compatible with observations. In particular, the limiting timescale for HERG lifetimes is equivalent to the time it takes for a high spinning retrograde accreting black hole to spin down at the Eddington limit. Based on the same simple evolutionary prescription for HERGs and LERGs in the model, we were also able to make contact with the observations of LERGs dominating in cluster environments as opposed to HERGs dominating in field environments. In this context, the model predicts the existence of short-lived higher redshift HERGs in clusters, which are the progenitors of the cluster LERGs. This redshift distribution is not predicted in the standard spin paradigm. We then showed how our evolutionary picture incorporates seemingly shutting down radio AGN that from the model perspective, instead, are understood as continuously accreting black holes through zero spin in ADAF mode. Although we have used the nomenclature of Tadhunter in our discussion of WLRG/FRII, we emphasize that the WLRG is modeled as a LERG or ADAF in its accretion state. By no means, in other words, are we suggesting new physics for this group of AGN. It is also worth pointing out that super-Eddington accretion has not been invoked in order to explain HERGs. Only near-Eddington accretion rates and lower are features of the applied model. We then showed how to understand/incorporate the redshift and environment distribution of mid-infrared emitting AGN as part of the HERG group of objects. This work constitutes a new series of applications of the model to observations and in particular it is the first time we recognize how the model can be applied to understand the distribution of radio galaxies in different environments. We hope to see AGN simulations eventually incorporate the differences in jet and disk feedback due to retrograde accretion to explore and go beyond our simple phenomenological framework.